\documentclass[9pt,twocolumn,twoside]{osajnl}

\pdfoutput=1
\usepackage{amsmath,amssymb}
\usepackage{subfig}
\usepackage{graphicx}
\usepackage{gensymb}

\newcommand{\be}{\begin{equation}}
\newcommand{\ee}{\end{equation}}
\newcommand{\bi}{\begin{itemize}}
	\newcommand{\ei}{\end{itemize}}
\newcommand{\bn}{\begin{enumerate}}
	\newcommand{\en}{\end{enumerate}}
\newcommand{\bea}{\begin{eqnarray}}
\newcommand{\eea}{\end{eqnarray}}

\newcommand{\siDiff}[1]{\ensuremath{{{x}}_{#1}}}

{}

\newcommand{\xb}{\mathbf{x}}
\newcommand{\ssb}{\mathbf{s}}
\newcommand{\zb}{\mathbf{z}}

\newcommand{\yb}{\mathbf{y}}
\newcommand{\wb}{\mathbf{n}}

\newcommand{\Hb}{\mathbf{H}}

\newcommand{\Cb}{\mathbf{C}}

\newcommand{\Ib}{\mathbf{I}}

\newcommand{\db}{\mathbf{d}}

\newcommand{\w}{\mathbf{w}}

\makeatletter
\newcommand{\longdash}[1][2em]{%
	\makebox[#1]{$\m@th\smash-\mkern-7mu\cleaders\hbox{$\mkern-2mu\smash-\mkern-2mu$}\hfill\mkern-7mu\smash-$}}
\makeatother
\newcommand{\omitskip}{\kern-\arraycolsep}

\journal{ol} 

\setboolean{shortarticle}{true}

\title{Compressive Spectral Imaging with Diffractive Lenses}

\author[1]{Oğuzhan Fatih Kar}
\author[1,*]{Figen S. Oktem}

\affil[1]{Department of Electrical and Electronics Engineering, Middle East Technical University (METU), Ankara, 06800, Turkey}

\affil[*]{Corresponding author: figeno@metu.edu.tr}

\begin{abstract}
Compressive spectral imaging enables to reconstruct the entire 
3D spectral cube from a few multiplexed images. Here, we develop a novel compressive spectral imaging technique using diffractive lenses. Our technique uses a coded aperture to spatially modulate the optical field from the scene and a diffractive lens such as a photon-sieve for both dispersion and focusing. 
Measurement diversity is achieved by changing the focusing behavior of the diffractive lens.
The 3D spectral cube is then reconstructed from 
highly compressed measurements taken with a monochrome detector. 
A fast sparse recovery method is developed to solve this large-scale inverse problem. The 
performance is illustrated at visible regime for various scenarios with different compression ratios 
through 
simulations. The results demonstrate that promising reconstruction performance can be achieved at high compression levels. This opens up new possibilities for high resolution spectral imaging with low-cost and simpler designs.
\end{abstract}

\setboolean{displaycopyright}{true}

\begin{document}

\maketitle

Spectral imaging 
is a fundamental diagnostic technique in physical sciences with application in diverse fields such as physics, chemistry, biology, medicine, astronomy, and remote sensing. Conventional techniques rely on a scanning process to build up the 3D spectral cube from a series of 2D measurements~\cite{okamoto1991}. One important disadvantage 
is that higher number of scans is needed with  increased spatial and spectral resolutions~\cite{cao2016computational}. This may lead to low light throughput, increased hardware complexity, and long acquisition times, resulting in temporal artifacts in dynamic scenes. 
Moreover, the temporal, spatial, and spectral resolutions are inherently limited 
as they 
are purely determined by the physical systems involved.

Compressive spectral imaging provides an effective way to overcome these limitations by passing on some of the burden to a computational system. It enables to reconstruct the entire spectral 
cube 
from a few 
multiplexed measurements via 
sparse recovery. This is made possible by compressive sensing (CS) which relies on two principles: sparsity of the spectral images in a 
transform domain and incoherence of the measurements. It is widely known that spectral images exhibit both spatial and spectral correlations, which allow sparse representations~\cite{cao2016computational}. For the incoherence of the measurements, different computational spectral imaging techniques have been proposed, as reviewed in~\cite{cao2016computational,oktem2018book}. Examples include coded aperture snapshot spectral imaging (CASSI) and its variants~\cite{wagadarikar2008,rajwade2013coded,salazar2019spectral,cao2016computational}, 
and compressive hyperspectral imaging by separable spectral-spatial operators~\cite{august2013}. 


In this letter, we develop a novel compressive spectral imaging technique 
named \emph{compressive spectral imaging with diffractive lenses} (CSID). CSID uses a coded aperture to spatially modulate the optical field from the scene and a diffractive 
lens such as a photon sieve~\cite{oktem2014icip,andersen2005} for both dispersion and focusing. The coded field is first passed through the diffractive lens and then 
recorded with a monochromatic detector. Measurement diversity 
is achieved by changing the focusing behavior of the diffractive lens.
A novel fast sparse recovery 
method is also developed to reconstruct the 
spectral cube from 
compressive measurements. The performance is  illustrated numerically for various settings.

Different than the earlier works that use diffractive lenses for spectral imaging~\cite{wang2018computational,hallada2017fresnel,nimmer2018spectral}, here we utilize them for the first time in a compressive modality.   
Moreover, our system performs dispersion and focusing with a single element (a diffractive lens) unlike conventional imaging spectrometers and computational spectral imaging systems like CTIS~\cite{okamoto1991} and CASSI~\cite{wagadarikar2008} for which collimating and focusing optics are also required in addition to a disperser (grating/prism).
Since diffractive lenses are also lightweight and low-cost to manufacture for a wide spectral range including x-rays and UV~\cite{attwood_book,oktem2014icip}, 
our approach enables high resolution spectral imaging with simpler and low-cost designs. 

\begin{figure}[t!]
	\begin{center} \vspace{-0.0in} 
		\includegraphics[width=3.2in]{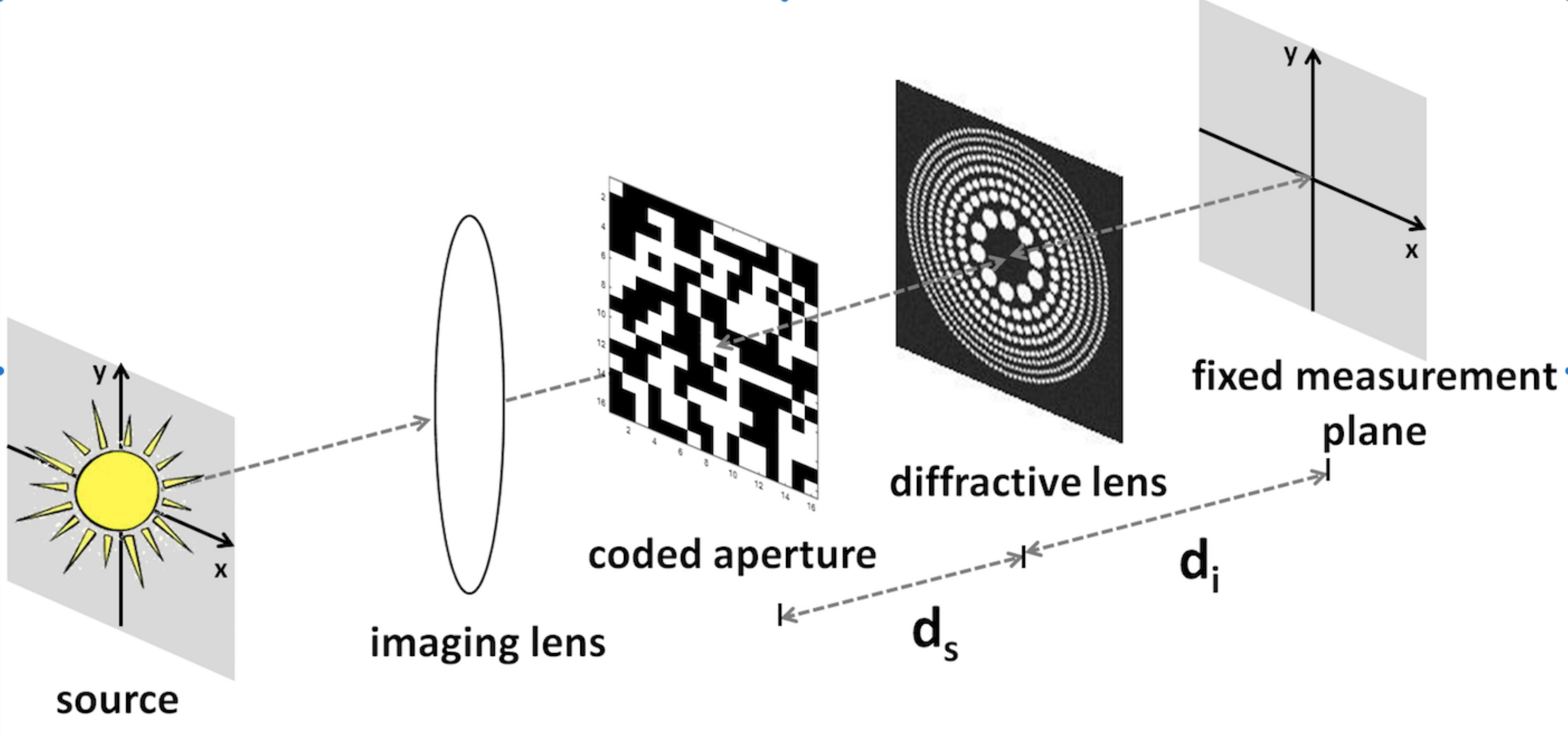}\hspace{0.00in} 
		\caption{Illustration of the CSID system.}
        \label{system}
		\vspace{-0.3in}
	\end{center} 
\end{figure}
Figure~\ref{system} illustrates the CSID system, which 
consists of (1) an imaging lens, (2) a coded mask, (3) a diffractive lens (such as a photon sieve), 
and (4) a monochrome detector~\cite{kar2018compressive}. First the image of the scene is formed on the plane of the coded mask, 
and then the 
coded field is 
passed through the diffractive lens.
Since the diffractive lens has a wavelength-dependent focal length, each spectral component
is exposed to a different amount of focus. As a result, each
measurement is 
a superposition of differently blurred and coded spectral bands. 
To achieve measurement diversity, a total of $K$ such measurements can be recorded 
by changing the focusing behavior of the diffractive lens.
Such few measurements can be obtained in different ways, such as with a programmable diffractive lens realized by a rapidly varying commercial spatial light modulator or digital micromirror device (DMD), 
or in a snapshot using multiple 
diffractive lenses and beam splitters.


The measurements obtained with the CSID system can be related
to the intensity of each spectral component as follows:
\vspace{-0.1in} 
\be
y_k(u,v)=\int \left( {f}_{\lambda}(u, v) \ast h_{\lambda,k}(u,v) \right) b(\lambda)d\lambda. \label{forwardModel1} \vspace{-0.05in}
\ee 
Here $y_k(u,v)$ represents the $k$th measurement, 
${f}_{\lambda}(u, v) = x_{\lambda}\left(-\frac{d_s}{d_i}  u, -\frac{d_s}{d_i} v\right) c_{\lambda}\left(-\frac{d_s}{d_i} u,-\frac{d_s}{d_i} v\right) $ is the coded and scaled intensity of the spectral field $x_{\lambda}(u,v)$ with 
coded aperture $c_{\lambda}(u,v)$. Assuming an ideal imaging lens with unit magnification, 
this coded and scaled intensity 
is convolved with the incoherent point-spread function (PSF) of the $k$th diffractive lens, $h_{\lambda,k}(u,v)$, which has a 
a closed-form expression given
elsewhere~\cite{oktem2018analytical}. Lastly, $b(\lambda)$ denotes the spectral response of the detector. 

We discretize the spectral field into $S$ spectral bands, and
$x_{s}(u,v)$ represents 
the intensity of the $s$th band with central wavelength $\lambda_s$. This spectral component is modulated with the coded mask 
pattern $c_s(u,v)$ at 
$\lambda_s$. 
The patterns $c_s(u,v)$ are the same for all wavelengths ($s=1,\hdots,S$) if an uncolored (
block-unblock) mask is used; however, 
these will be different if a colored coded mask~\cite{cao2016computational}
is used instead. 
The coded aperture $c_s(u,v) = \sum_{m,n} c_s[m,n] {\rm{rect}}(\frac{u}{\Delta_c}-{m},\frac{v}{\Delta_c}-{n})$ is a pixelated array with a pixel size of $\Delta_c$, and $c_s[m,n]$ denotes the value of the coded aperture at pixel $(m,n)$. 

After discretizing the 
field along the spectral dimension, discretization along the spatial dimensions is also needed to arrive at a discrete model. 
Replacing each spatially continuous function with its discretized version, we obtain the following model:
\vspace{-0.1in} 
\be
y_k[m,n]=\sum_{s=1}^S ( \, b_s \, \siDiff{s}[m, n] c_s[m,n]) \ast h_{\lambda_s,k}[m, n]. \label{forwardModel2} \vspace{-0.05in}
\ee 
Here, $y_k[m,n]$ denotes the $k$th measurement obtained over $N_x\times N_y$ detector pixels, and corresponds to the samples of $y_k(u,v)$, i.e. $y_k[m,n]=y_k(m\Delta,n\Delta)$.
The sampling interval $\Delta$ 
is equal to the pixel size of the detector. The coded aperture pixel size can be chosen as an integer multiple of $\Delta$ to avoid the need for subpixel positioning accuracy. Here, we choose $\Delta_c = \Delta$ for simplicity. 
Moreover, $\siDiff{s}[m, n]$ and $h_{\lambda_s,k}[m, n] $ are the uniformly sampled versions of their continuous counterparts with the same sampling interval $\Delta$. 
Lastly, $b_s$ represents the coefficient resulting from the response of the detector at the central wavelength $\lambda_s$.


This discrete 
model can be expressed in the following 
form: 
\be
{\yb}={\Hb}{\Cb}{{\xb}}+{\wb} , \label{forwardModelc}
\ee
where $\yb = [ \yb_1^T, ..., \yb_K^T ]^T \in {\mathbb{R}}^{KN}$ is vertically concatenated measurement vector with $N\triangleq N_x N_y$ and ${\yb}_k\in {\mathbb{R}}^{N}$ denoting the $k$th measurement vector. Similarly, $\xb = [ \xb_1^T, ..., \xb_S^T ]^T  \in {\mathbb{R}}^{SN}$ is the concatenated image vector with $\xb_s\in {\mathbb{R}}^{N}$ denoting the spectral image vector 
at wavelength $\lambda_s$.
The $K N \times S N$ matrix $\Hb$ consists of $N \times N$ convolution matrices representing the convolutions with PSFs $h_{\lambda_s,k}[m, n]$. The diagonal matrix ${\Cb}\in {\mathbb{R}}^{SN\times SN}$ 
performs the overall coding operation, and has values $0$ or $1$ along its diagonal. Finally, the vector $\wb=[{\bf n}_1^T, ... ,{\bf n}_K^T]^T$ denotes the 
noise, which is often white Gaussian. 
In our setting, the number of measurements ($K$) is smaller than the number of spectral bands ($S$), which results in an under-determined system. 

In the inverse problem, the goal is to reconstruct the unknown spectral images, $\xb$, from their compressive superimposed measurements, ${\yb}$, which 
contain their coded and blurred versions. This problem is inherently ill-posed. There are a variety of approaches to solve such ill-posed linear inverse problems. Here, to exploit the sparsity of the spectral images after some transformation $\Phi$, we formulate the inverse problem as the following constrained optimization problem:
\be
\min_{{{\xb}}} \;  \Vert \Phi\xb\Vert_1~ {\text{subject to}}~ ||  {\yb} - {\Hb}{\Cb}{\xb} ||_2\leq\epsilon,
\label{inverseProblem2} \vspace{-0.0in}
\ee 
where $\epsilon\geq0$ is a parameter that depends on noise variance. 
Here $\ell_1$-norm enforces the 
sparsity of the spectral cube after transformation with $\Phi$, as motivated by the CS theory.

To solve the resulting optimization problem, we convert our constrained problem to an unconstrained problem by adding the constraint to the objective function as a penalty function:
\be
\min_{{{\xb}}} \;  \Vert \Phi\xb\Vert_1+\iota_{(|| {\yb} - {\Hb}{\Cb}{{\xb}} ||_2\leq\epsilon)}(\xb),
\label{inverseProblem3} \vspace{-0.0in}
\ee 
where the indicator function $\iota_{(|| {\yb} - {\Hb}{\Cb}{{\xb}} ||_2\leq\epsilon)}(\xb)$ takes value 0 if the constraint is satisfied, and $+\infty$ otherwise.
We solve this problem by developing a fast reconstruction algorithm that is based on alternating direction method of multipliers (ADMM)~\cite{c-salsa}. After variable-splitting, 
we arrive at the following problem:
\begin{align}
\begin{array}{cc}
\underset{\xb,\zb^{(1)},\zb^{(2)}}{\text{minimize}} & \Vert \Phi\zb^{(1)}\Vert_1 +\iota_{(|| {\yb} - {\zb^{(2)}} ||_2\leq\epsilon)}(\zb^{(2)}) \\
\text{subject to} & \zb^{(1)} = \xb ,~~ \zb^{(2)} = {\Hb}{\Cb}\xb
\end{array}  \label{eq:admm_}
\end{align} 
where $\zb^{(1)}$, $\zb^{(2)}$ are the auxiliary variables in the ADMM framework. After expressing the problem in \eqref{eq:admm_} in augmented Lagrangian form~\cite{c-salsa}, minimization over $\xb$, $\zb^{(1)}$, and $\zb^{(2)}$ is needed. Here, we minimize over each in an alternating fashion.

For minimization over $\xb$, we face a least-squares problem which has the following normal equation:
\begin{align}
 (\Ib + {\Cb}{\Hb}^{H}{\Hb}{\Cb})\xb_{k+1} =(\zb^{(1)}+\db^{(1)}+{\Cb}{\Hb}^{H}(\zb^{(2)}+\db^{(2)}))
\label{x-ls_final}
\end{align}
with $\db$ denoting the dual variable in the ADMM framework. A  direct matrix inversion approach for solving the linear system in \eqref{x-ls_final} is not feasible for large-scale spectral cubes. Here, we solve this iteratively using the conjugate-gradient method. For this iterative process, forming any of the matrices is not required, which provides huge savings for the memory 
and 
computation time. Specifically, multiplications with matrices ${\Hb}$ and ${\Hb}^H$ correspond to summation of some convolutions. That is, for multiplication with ${\Hb}$ matrix, we simply take 2D
Fourier transforms of underlying PSFs $h_{\lambda_1,k}[m, n],\hdots, h_{\lambda_S,k}[m, n]$ and the spectral images $\siDiff{1}[m, n],\hdots,\siDiff{S}[m, n]$, multiply them element-wise, and then sum. 
For multiplication with ${\Hb}^H$ matrix, a similar operation is performed using the PSFs $h_{\lambda_s,1}[m, n],\hdots, h_{\lambda_s,K}[m, n]$. 
Moreover, the 
multiplication with ${\Cb}$ 
corresponds to simple element-wise multiplications with coded apertures 
$c_s[m,n]$.



Minimization over $\zb^{(1)}$ 
requires the following operation:
%
\begin{align}
\begin{array}{cc}
 \zb^{(1)}_{k+1}=\Phi^{-1}(\textit{soft}(\Phi(\xb_{k+1}-\db^{(1)}_k),\frac{1}{\mu})), 
 \end{array}  \label{eq:normaleq_prox}
\end{align} 
Here, $\textit{soft}(\w,\tau)$ denotes the soft-thresholding operation and is component-wise computed as $\w_i\rightarrow \textit{sign}(\w_i)\max(|\w_i|-\tau,0)$ for all $i$, with $\textit{sign}(\w_i)$ taking value $1$ if $w_i>0$ and $-1$ otherwise~\cite{c-salsa}.
That is, the solution in \eqref{eq:normaleq_prox} can be obtained through transformation with $\Phi$, followed by soft-thresholding with parameter $1/\mu$, and inverse transformation operation $\Phi^{-1}$. 

For minimization over $\zb^{(2)}$,
a projection of $\ssb\triangleq({\Hb}{\Cb}\xb_{k+1}-\db^{(2)}_k) $ onto $\epsilon$-radius hypersphere centered at $\yb$ is required~\cite{c-salsa}. This projection has the following form: 
\begin{align}
\begin{array}{cc}
\zb^{(2)}_{k+1}=
\begin{cases}
\yb + \epsilon\frac{\ssb-\yb}{\Vert \ssb-\yb \Vert_2},~ &\text{if}~~ \Vert \ssb-\yb \Vert_2 > \epsilon \\
\ssb,~ &\text{if}~~ \Vert \ssb-\yb \Vert_2 \leq \epsilon.
\end{cases}
 \end{array}  \label{eq:normaleq_prox_proj}
\end{align} 

As a result, we have three update steps resulting from the ADMM formulation, i.e. $\xb$-update, $\zb^{(1)}$-update, and $\zb^{(2)}$-update. The overall algorithm is summarized in Table~\ref{tab:pssi}.

\begin{table}[h]\vspace{-0.2in}
	\centering
	\caption{Reconstruction algorithm for CSID} 
	\label{tab:pssi}
	\begin{tabular}{l}
		\hline
		\vspace{-0.05in}
		\textbf{Input:} Compressive measurements $\yb$ as in \eqref{forwardModelc}. \\
		\vspace{-0.05in}
		\textbf{Initialization:} 
		Iteration count $k = 0$, choose $\mu>0$, $\epsilon$,\\
		\vspace{-0.05in}
		$\zb_0^{(1)}$, $\zb_0^{(2)}$,$\db_0^{(1)}$, $\db_0^{(2)}$. \\
		\vspace{-0.05in}
		\textbf{Main Iteration:} 
		Repeat until stopping criterion satisfied. \\
		\vspace{-0.05in}
		1. Calculate spectral images $\xb_{k+1}$ by solving \eqref{x-ls_final} \\
		\vspace{-0.05in}
		using conjugate-gradient algorithm. \\
		\vspace{-0.00in}
		2. Calculate $\zb^{(1)}_{k+1}$ using soft-thresholding in \eqref{eq:normaleq_prox}. \\
		\vspace{-0.00in}
		3. Calculate $\zb^{(2)}_{k+1}$ using projection in \eqref{eq:normaleq_prox_proj}.\\
		~4. Update $\db^{(1)}_{k+1}$ as ${\db}^{(1)}_{k+1} = {\db}^{(1)}_k - (\xb_{k+1}-\zb^{(1)}_{k+1})$. \\
		\vspace{-0.05in}
		5. Update $\db^{(2)}_{k+1}$ as ${\db}^{(2)}_{k+1} = {\db}^{(2)}_k -({\Hb}{\Cb}\xb_{k+1}-\zb^{(2)}_{k+1})$. \\
		\vspace{-0.02in}
		\textbf{Output:} Spectral images $\xb$. \\
		\hline
	\end{tabular} \vspace{-0.1in}
\end{table}

We now present numerical simulations to illustrate the performance 
and compare with 
CASSI. 
We consider a 
spectral dataset of size $820 \times 820 \times 31$ ($31$ wavelengths 
between $410-710$ nm with $10$ nm interval), 
taken from an online hyperspectral 
database~\cite{nascimento2002statistics} and referred as Objects data in a CASSI work~\cite{rajwade2013coded}. 
For this dataset, $25$ outlier voxel values are dropped to $0.3$, and 
the spectral cube is scaled to 
$[0,~1]$. 
As the diffractive lens, 
photon sieves are used with a smallest hole diameter of $\delta=8$ $\mu$m, providing Abbe's spatial resolution of $8$ $\mu$m. 
Pixel size of the detector, $\Delta$, is chosen as $4$ $\mu$m 
to match 
this spatial resolution.

Measurements are taken by changing the focusing behavior of the photon sieve. 
For each measurement, the detector to diffractive lens distance is fixed as $f_0=2.56$ cm, and 
the design is changed to focus a different wavelength at this distance. For this purpose, the outer diameter of the 
sieve is changed as $D_k = \Tilde{\lambda}_k f_0/\delta$~\cite{attwood_book},
where $\Tilde{\lambda}_k$ is the wavelength focused at the $k$th measurement. For example, if 
$\Tilde{\lambda}_k=560$ nm, 
the diameter $D_k = 1.8$ mm.
Moreover, the expected spectral resolution is 
$4 \, \delta^2 /f_0=10$ nm, as given by the spectral bandwidth of the diffractive lens~\cite{attwood_book}[Chap.~9]. Note that this expected spectral resolution matches to the spectral sampling interval, i.e. $10$ nm. 

\begin{figure}[th!b]
	\begin{center} \vspace{-0.00in} 
		\includegraphics[width=2.5in]{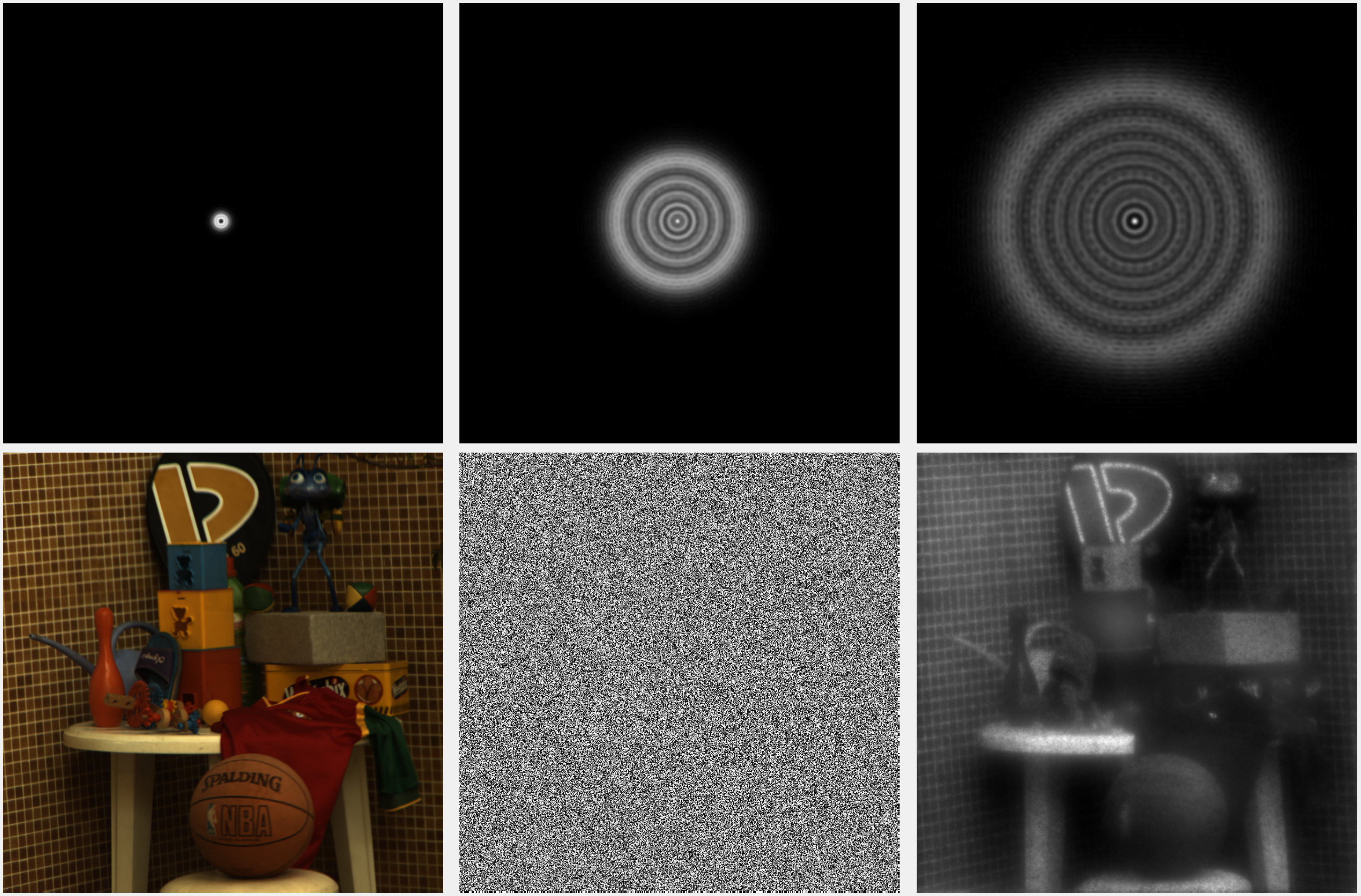}
        \hspace{0.00in} 
       \vspace{-0.0in}
		\caption{Demonstration of compressive measurements 
		for $K=3$ case. In this measurement, $430$ nm is focused onto the detector plane, while all other spectral components are defocused. Top row: PSFs of the photon-sieve for spectral components 
		at $410$ nm, $560$ nm, and $710$ nm.  
		Bottom row: Superimposed true image, sample mask pattern, sample compressive measurement.
		}
		\vspace{-0.10in}
        \label{fig:meas_etc}
	\end{center} 
    \end{figure}

The 
compressive measurements are simulated using the 
model in \eqref{forwardModelc} with additive Gaussian noise. In each measurement, the system applies the same masking operation to each spectral band using a traditional block-unblock
mask, whose entries are drawn from a Bernoulli distribution as shown in Fig.~\ref{fig:meas_etc}.
After the coded field passes through the photon sieve, we capture measurements at the same plane by changing the 
outer diameter of the sieve. 
A sample compressive measurement 
is shown in Fig.~\ref{fig:meas_etc} 
together with the true spectral cube superimposed 
along the spectral dimension. In this measurement, $430$ nm is focused by the sieve onto the detector plane, while all other spectral components are defocused. 
To illustrate 
this, we also provide the acting PSFs for 
three spectral components,
which show the different amount of blur. 
As seen, the measurements involve not only the superposition of all spectral bands but also significant amount of blur and degradation. 

\begin{table}[h!] \vspace{-0.2in}
	\renewcommand{\arraystretch}{1.3}
	\fontsize{7}{9}\selectfont
	\centering
	\caption{Comparison of 
	reconstruction PSNRs (dB) / SSIMs / SAMs for different compressive 
	scenarios and SNRs.}
	\label{Table:method_comparison}
	\vspace{-0.0in}
	\begin{tabular}{|c|c|c|c|c|}
		\hline
		\textbf{SNR (dB)} & \textbf{$K=2$} & \textbf{$K=3$} & \textbf{$K=4$} \\
		\hline
		22& 28.62/0.73/19.3$\degree$&32.13/0.84/12.4$\degree$ & 32.60/0.85/11.8$\degree$    \\
		\hline
		28&  28.93/0.73/18.9$\degree$&32.73/0.85/11.8$\degree$ & 33.42/0.87/10.9$\degree$   \\
		\hline
		34&  29.23/0.73/18.7$\degree$&33.16/0.86/11.4$\degree$ & 34.19/0.88/10.2$\degree$   \\
		\hline	
	\end{tabular} \vspace{-0.10in}
\end{table}

\begin{figure*}[h!]
	\begin{center} \vspace{-0.0in} 
		\includegraphics[width=5.7in]{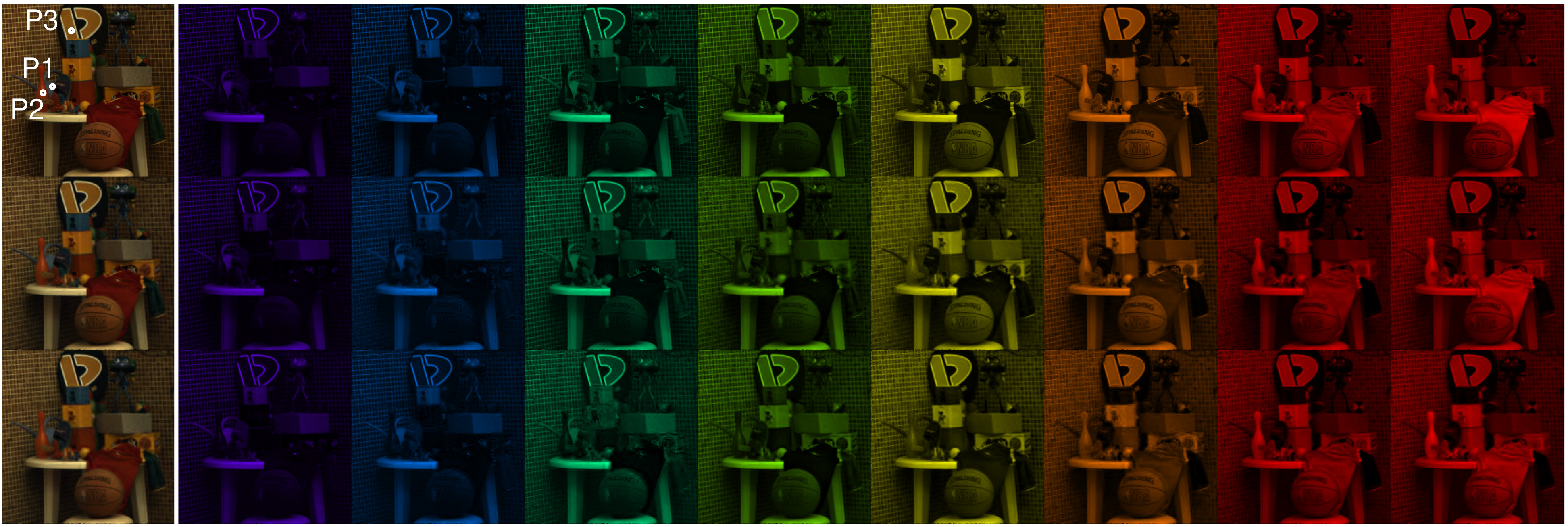}\hspace{0.00in}
		\vspace{-0.10in}
  		\caption{Sample reconstructed 
			images 
			from compressive measurements when 
			SNR$=28$ dB. In the left of each row, superimposed 
			spectral cube along the spectral dimension
			is shown; other columns contain 
			spectral images at wavelengths $420$, $460$, $500$, $540$, $580$, $620$, $660$, and $700$ nm. Top to bottom: true images, 
			reconstructions with $K=4$ and $K=3$ measurements.}\vspace{-0.1in}
			\label{fig:recons_res} 
	\end{center} 
\end{figure*}
\begin{figure}[h!]
	\begin{center} \vspace{-0.2in} 

		\includegraphics[width=3.4in]{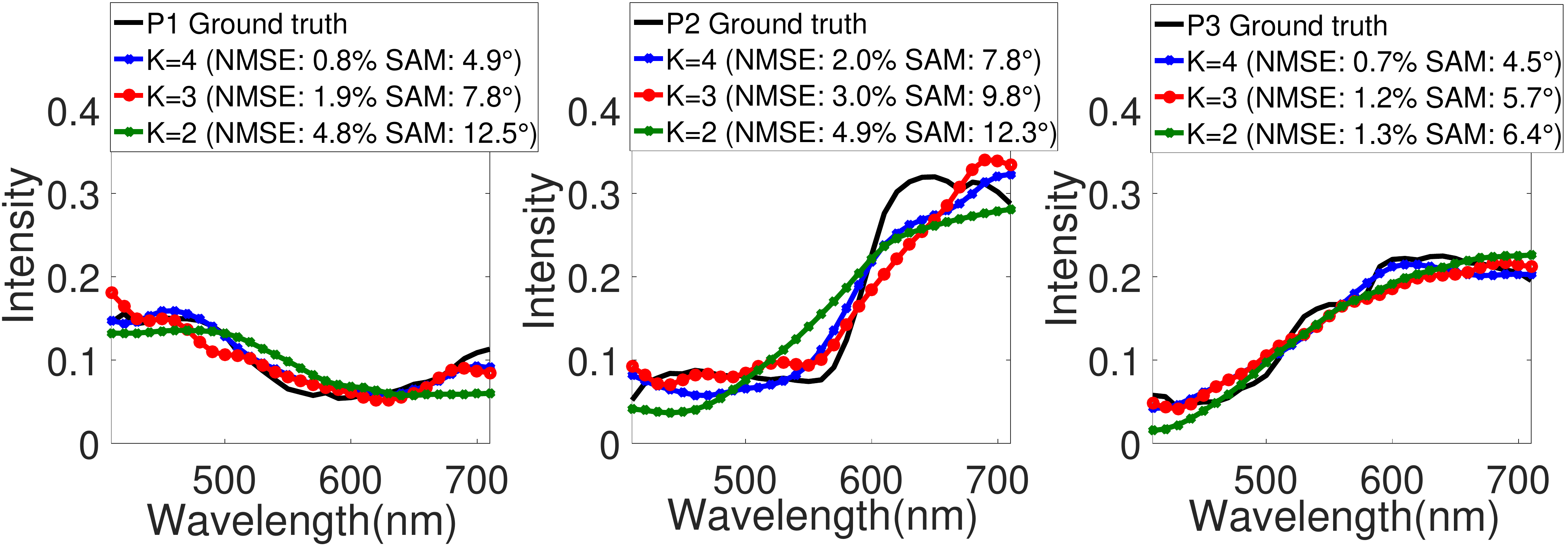} \vspace{-0.10in}
		\caption{Spectrum at the points P1, P2, P3 for 
		SNR$=28$ dB. 
		}
		\vspace{-0.3in} 
		\label{fig:compressions}
	\end{center} 
\end{figure}

We consider different 
compressive scenarios with $K=2$, $3$ and $4$ measurements. For each case, 
equidistant wavelengths from the spectral range $410$-$710$ nm are chosen to be focused onto the detector plane.
More specifically, the chosen wavelengths are $\{500,~610\}$~nm for $K=2$,
$\{430,~560,~680\}$~nm for $K=3$, 
and $\{420,~510,~600,~690\}$ nm for $K=4$. 
These cases with $K=2,3,4$ correspond to compression levels (CLs), $100 \times \left( 1 - K/S \right)$, of 
$93.5\%$, $90.3\%$ and $87.1\%$,
respectively. These are equivalent to reconstructing the spectral cube from 
$6.5\%$, $9.7\%$ and $12.9\%$ 
data.

To analyze medium to low noise cases, input SNRs of $22$, $28$, and $34$ dB are considered, 
by adding Gaussian noise with standard deviation equal to $1 \%$, $0.50 \%$, and $0.25 \%$ of the maximum value in the noiseless measurements. 
Reconstructions are obtained from these compressive 
noisy measurements using the algorithm in Table~\ref{tab:pssi}. Similar to previous compressive spectral imaging approaches~\cite{cao2016computational}, we enforce sparsity in a Kronecker basis $\Phi=\Phi_1\otimes\Phi_2$ where $\Phi_1$ is the basis for 2D Symmlet-8 wavelet and $\Phi_2$ is the 1D discrete cosine (DCT) basis. This transformation 
is computed by first taking the 2D Symmlet-8 transform of each spectral image and then 1D DCT along the spectral dimension.

The average reconstruction performance for all cases is given in Table~\ref{Table:method_comparison} in terms of PSNR, SSIM, and spectral angular mapper (SAM). As seen, PSNR is above $28.5$ dB, SSIM is above $0.73$, and SAM is less than 19.3$\degree$ for all cases, which demonstrates 
faithful reconstruction 
even at high compression levels. 
Moreover, for $K=3$ case (i.e. 
reconstruction from $9.7\%$ data), PSNR is greater than $32.13$ dB and SSIM is greater than $0.84$ for all three SNRs. 
These values are 
better than the multi-frame CASSI results for the same dataset and compression level given in~\cite{rajwade2013coded} (PSNR=$27.04$ dB, SSIM=$0.82$). In addition, the performance degrades gracefully with decreased 
SNR and increased compression. 

To visually evaluate the results, we provide in Fig.~\ref{fig:recons_res} 
the reconstructed spectral images at different compression levels for SNR=$28$ dB, together with the true 
images. 
As seen, the image details and edges, as well as the spectral variations, are well preserved in the reconstructions. 
In the left of each row, superimposed 
spectral cube is also shown, which is 
similar to the true one. Hence, the results demonstrate successful reconstruction of the spectral 
cube 
at compression levels 
as high as $\sim 90\%$.

To also demonstrate the successful recovery along the spectral dimension, we select three representative points with different spectral characteristics, as shown as P1, P2, and P3 in Fig.~\ref{fig:recons_res}. 
The reconstructed spectra at these points are plotted in Fig.~\ref{fig:compressions}, 
together with the ground truth. 
As seen, the spectrum is recovered  successfully
at all compression ratios for each point. 
To numerically evaluate the spectrum recovery, 
SAM and percentage 
mean squared error (NMSE) 
values are also given in the legends.


In these 
results, the pixel size of the detector 
and the reconstruction grid are chosen to match the expected spatial resolution of the diffractive lens. 
Because the developed 
modality is based on computational imaging 
and compression is performed along the spectral direction, effective spectral resolution not only depends on the 
diffractive lens design, but also on the scene content (i.e. the spectral correlation). Although the imaging performance appears to be robust to higher compression levels and noise, clearly increasing the number of measurements improves the reconstructions. However, this comes with the cost of increased acquisition time (i.e. 
undesirable for dynamic scenes). 

In summary, we have presented a novel compressive spectral imaging modality 
with a simple optical configuration involving a coded aperture and 
a diffractive lens. Together with the developed reconstruction algorithm, promising imaging 
performance is achieved even at high compression levels.
Since the system 
performs compression along the spectral dimension, successful reconstructions can be obtained for spectrally-correlated scenes.
Although the presented results are for the visible range, the 
imaging concept is equally applicable to other regimes as well. 
Moreover, the part of the 
imaging system after the coded aperture is shift-invariant unlike earlier systems. This enables easier design, faster reconstruction, and simpler calibration. In particular, for calibration, measuring the PSFs is sufficient, instead of the system response for each voxel.
The performance can be further improved with the 
use of colored coded apertures. Future work will focus on the experimental demonstration.


Different than the earlier compressive spectral imagers that rely on prisms/gratings to disperse the optical field and require additional collimating/re-imaging optics, 
we use a single diffractive lens
to 
achieve both dispersion and focusing. Moreover, unlike conventional collimating/imaging optics,
diffractive lenses are lightweight and low-cost to manufacture for a wide spectral range including x-rays and UV. Hence 
this work
opens up new possibilities for high resolution spectral imaging with low-cost and simpler designs
in a wide range of applications.


\textbf{Funding:}
Scientific and Technological Research Council of Turkey (TUBITAK), 3501 Research Program, 117E160.



\ifthenelse{\equal{\journalref}{aop}}{%
\section*{Author Biographies}
\begingroup
\setlength\intextsep{0pt}
\begin{minipage}[t][6.3cm][t]{1.0\textwidth} 
  \begin{wrapfigure}{L}{0.25\textwidth}
    \includegraphics[width=0.25\textwidth]{john_smith.eps}
  \end{wrapfigure}
  \noindent
  {\bfseries John Smith} received his BSc (Mathematics) in 2000 from The University of Maryland. His research interests include lasers and optics.
\end{minipage}
\begin{minipage}{1.0\textwidth}
  \begin{wrapfigure}{L}{0.25\textwidth}
    \includegraphics[width=0.25\textwidth]{alice_smith.eps}
  \end{wrapfigure}
  \noindent
  {\bfseries Alice Smith} also received her BSc (Mathematics) in 2000 from The University of Maryland. Her research interests also include lasers and optics.
\end{minipage}
\endgroup
}{}

\end{document}